\documentclass[letterpaper]{article} 
\usepackage{aaai24}  
\usepackage{times}  
\usepackage{helvet}  
\usepackage{courier}  
\usepackage[hyphens]{url}  
\usepackage{graphicx} 
\usepackage{natbib}  
\usepackage{caption} 
\usepackage{algorithm}
\usepackage{algorithmic}

\usepackage{amsmath}
\usepackage{amsfonts}

\usepackage{newfloat}
\usepackage{listings}

\DeclareCaptionStyle{ruled}{labelfont=normalfont,labelsep=colon,strut=off} 
\floatstyle{ruled}
\newfloat{listing}{tb}{lst}{}
\floatname{listing}{Listing}
\title{Music Style Transfer with Time-Varying Inversion of Diffusion Models}
\author {
    Sifei Li\textsuperscript{\rm 1,\rm 2},
    Yuxin Zhang\textsuperscript{\rm 1,\rm 2},
    Fan Tang\textsuperscript{\rm 3},
    Chongyang Ma\textsuperscript{\rm 4},
    Weiming Dong\textsuperscript{\rm 1,\rm 2}\thanks{Corresponding author},
    Changsheng Xu\textsuperscript{\rm 1,\rm 2}
}
\affiliations {
    \textsuperscript{\rm 1}MAIS, Institute of Automation, Chinese Academy of Sciences \\
    \textsuperscript{\rm 2}School of Artificial Intelligence, University of Chinese Academy of Sciences\\
    \textsuperscript{\rm 3}Institute of Computing Technology, Chinese Academy of Sciences \\
    \textsuperscript{\rm 4}Kuaishou Technology\\
    \{lisifei2022, weiming.dong, changsheng.xu\}@ia.ac.cn, tangfan@ict.ac.cn, chongyangm@gmail.com
}

\usepackage{bibentry}

\begin{document}

\maketitle
\begin{figure*}[t]
\centering
   \includegraphics[width=\linewidth]{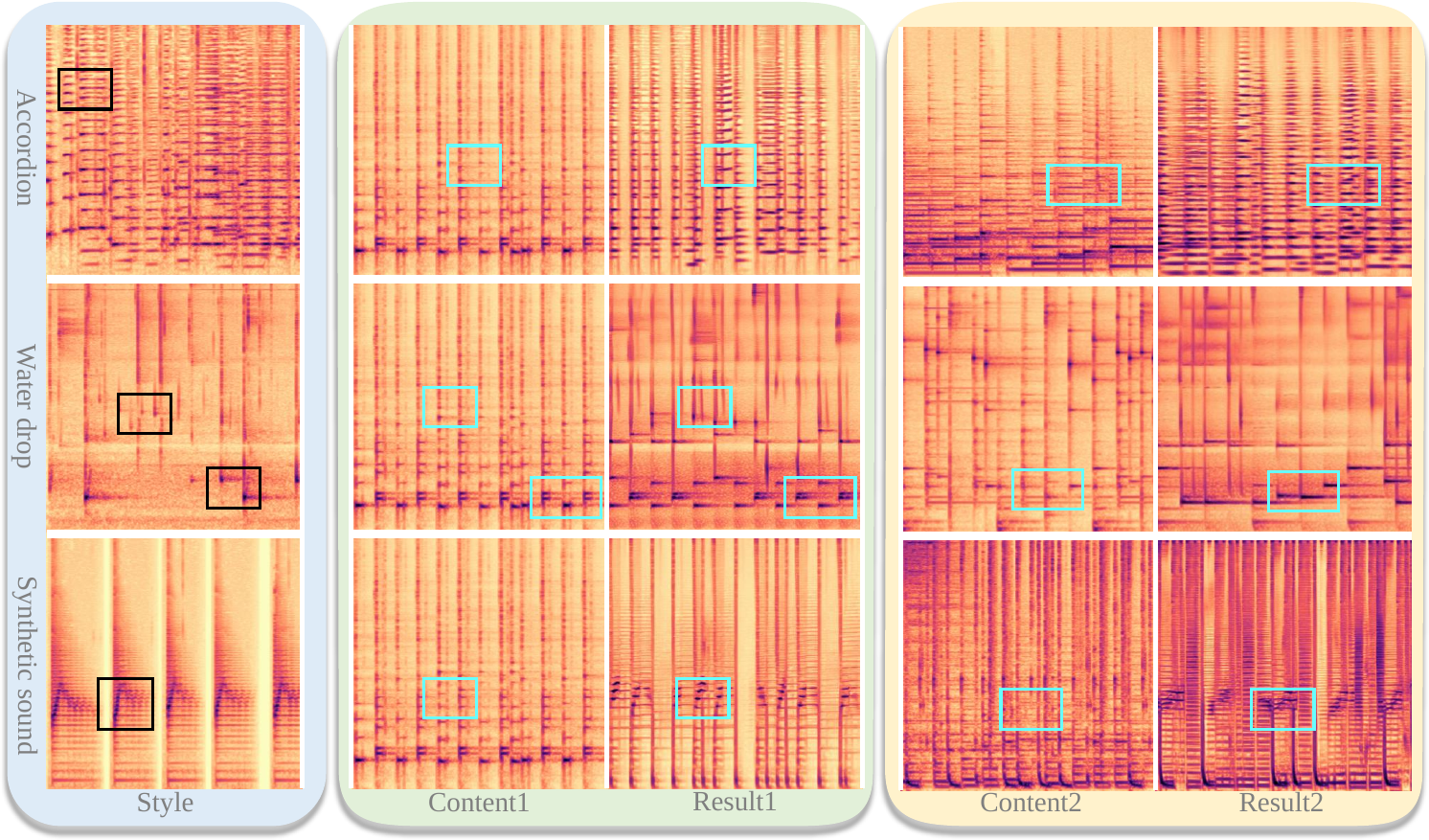}
   \caption{Music style transfer results using our method. Our approach can accurately transfer the style of various mel-spectrograms (e.g., instruments, natural sounds, synthetic sound) to content mel-spectrograms using minimal reference data, even as little as a five-second clip. In the style mel-spectrograms, the black box highlights the regions with prominent texture. It can be observed in the blue boxes that the style transfer results preserve a similar structure to the content mel-spectrograms while exhibiting similar texture to the style mel-spectrograms.}
\label{fig:result}
\end{figure*}
\begin{abstract}
With the development of diffusion models, text-guided image style transfer has demonstrated high-quality controllable synthesis results. However, the utilization of text for diverse music style transfer poses significant challenges, primarily due to the limited availability of matched audio-text datasets. Music, being an abstract and complex art form, exhibits variations and intricacies even within the same genre, thereby making accurate textual descriptions challenging. This paper presents a music style transfer approach that effectively captures musical attributes using minimal data. 
We introduce a novel time-varying textual inversion module to precisely capture mel-spectrogram features at different levels.
During inference, we propose a bias-reduced stylization technique to obtain stable results. Experimental results demonstrate that our method can transfer the style of specific instruments, as well as incorporate natural sounds to compose melodies. Samples and source code are available at https://lsfhuihuiff.github.io/MusicTI/.
\end{abstract}
\section*{Introduction}
If a picture is worth a thousand words, then every melody is timeless.
Music is an essential art form in human society, and a change in music style can offer listeners a completely new experience and perception.
For a long time, music creation has had high barriers to entry. However, music style transfer has opened up possibilities for ordinary individuals to achieve personalized music experiences. Music style transfer refers to the process of transferring the style of a given audio clip to another without altering its melody.
Sound is omnipresent in our lives, so inspired by music creators who utilize natural sounds in their compositions\footnote{How natural sounds can be involved in music production is well explained by \url{https://youtu.be/ixiiesRtgKU?list=RDixiiesRtgKU}; \url{https://theworld.org/stories/2021-03-14/nature-always-singing-now-you-can-make-your-own-music-natures-sounds}.}, music style transfer can be extended to encompass various types of sound examples.
 
Deep learning-based music style transfer has been a hot research topic in recent years.
Some works~\cite{alinoori2022music,choi2023pop2piano} can stylize music with a specific timbre to a specific or a few instruments, while others~\cite{huang2019timbretron,chang2021semi, bonnici2022timbre, wu2023transplayer} have achieved many-to-many music style transfer but restrict the transformation to a finite set of styles presented in the training data.
There are efforts~\cite{cifka2020groove2groove, cifka2021self} to explore one-shot music style transfer, but they still have difficulties in handling natural sounds. With the development of large language models, some
works~\cite{web_reference, liu2023audioldm, schneider2023mo, huang2023noise2music} explore text-guided music generation and demonstrate remarkable capacity for generating impressive results.
Specially, MusicLM~\cite{agostinelli2023musiclm} and MUSICGEN~\cite{copet2023simple} implement music style transfer by conditioning on both textual and melodic representations. However, existing methods can only achieve common style transfer based on coarse descriptions of genres (e.g., ``rock'', ``jazz''), instruments (e.g., ``piano'', ``guitar'', ``violin''), or performance forms (e.g., ``chorus'', ``string quartet''). They lack the ability to handle niche instruments such as cornet or erhu. Furthermore, these methods are insufficient to address complex scenarios involving the description of natural sounds or synthesized audio effects.

To alleviate all the above problems and leverage the generative capabilities of pretrained large-scale models, we propose a novel example-guided music stylization method. Our approach aims to achieve music style transfer based on arbitrary examples, encompassing instruments, natural sounds, and synthesized sound effects. Given an audio clip, we can transfer its style to arbitrary input music which is used as content.
As illustrated in Figure~\ref{fig:result}, our method can transfer the texture of the style mel-spectrograms to the local regions of the content mel-spectrograms, while preserving the structure of the content mel-spectrograms.
 
To achieve this goal, we seek to obtain an effective style representation of the input audio. Inspired by Textual Inversion~\cite{gal2022image}, which utilizes a pseudo-word to represent a  specific concept through the reconstruction of target images, we aim to learn a pseudo-word that represents the style audio in a similar manner.
However, we expect to avoid introducing the content of the style audio during the stylization process.
We suppose that different timesteps of the diffusion model focus on different levels of features. Therefore, we propose a time-varying textual inversion module, where the emphasis of text embedding shifts from texture to structure of the style mel-spectrogram as the timestep increases.
Futhermore, we use a partially noisy mel-spectrogram of the content music as the content guidance. As a result, when using the pseudo-word as guidance in the execution of DDIM~\cite{song2020denoising}, it becomes a partial denoising process. This scheme naturally excludes structure-related timesteps, which are associated with melody or rhythm, from participating in the stylization process.
Meanwhile, it preserves the melody or rhythm of the content mel-spectrogram. To reduce bias of diffusion models on content preservation, we add noise to the mel-spectrogram using the predicted noise instead of random noise, resulting in a more stable stylization result.

Our contributions can be summarized as follows:
\begin{itemize}
    \item We propose a novel example-based method for music style transfer with
    time-varying textual inversion.
    \item Our approach enables the use of non-musical audio for music style transfer and achieves highly creative results.
    \item Experimental results demonstrate that our method outperforms existing approaches in both qualitative and quantitative evaluations.
\end{itemize}
\section*{Related Work}

\paragraph{Music style transfer.}
Deep learning-based music style transfer has been widely studied as a typical mechanism of music generation.
\citet{dai2018music} explores the concept of music style transfer and analyzes its development. Many works have conducted further research on music style transfer using various deep learning frameworks~\cite{grinstein2018audio, bitton2018modulated, mor2019universal, huang2019timbretron, lu2018transferring, brunner2018midi, lu2019play, jain2020att}.
TimbreTron~\cite{huang2019timbretron} employs image style transfer techniques to achieve timbre transfer across multiple styles.
\citet{grinstein2018audio} explore timbre transfer between arbitrary audios based on CNN-extracted statistical features of audio styles.
Groove2Groove~\cite{cifka2020groove2groove} adopts an encoder-decoder structure to achieve one-shot style transfer for symbolic music.
\citet{cifka2021self} employs vector-quantized variational autoencoder (VQ-VAE) for one-shot music style transfer without being restricted to the training data, yielding good performance even on real-world data.
Music-STAR~\cite{alinoori2022music} explores style transfer between multi-track pieces, but it is limited to specific instruments.
\citet{bonnici2022timbre} utilize variational autoencoders (VAE) with generative adversarial networks for timbre transfer in both speakers and instruments.
Pop2Piano~\cite{choi2023pop2piano} uses transformer architecture to achieve the transformation from popular music to piano covers.
\citet{chang2021semi} and \citet{wu2023transplayer} implement many-to-many timbre transfer using autoencoders. However, these methods are seriously limited by the training data for achieving satisfactory timbre transfer results.
\citet{Wu:MuseMorphose:2023} combines Transformers and VAE to create a single model that can generate music with both long sequence modeling capability and user control over specific parts.
Above methods can generate good music style transfer results, but they can only achieve single-style transfer or require a large amount of training data, while failing to generate high-quality music with natural sound sources.

\paragraph{Text-to-music generation.}
Large-scale multimodal generative modeling has created milestones in text-to-music generation.
Make-An-Audio ~\cite{huang2023make} utilizes a prompt-enhanced diffusion model to implement audio representation generation in the latent space. AudioLDM~\cite{liu2023audioldm} uses Latent Diffusion Model (LDM) and CLAP~\cite{wu2023large} to generate audio (including music), and is the first work that can perform zero-shot text-guided audio editing.
Tango~\cite{ghosal2023tango} achieves high performance on text-to-audio task with limited data by utilizing the training concept of InstructGPT~\cite{ouyang2022training}.
However, the above works tend to focus on various sounds in the natural world, and their ability to generate music is limited.
Recently, diffusion models and transformers have gained significant popularity in the realm of music generation. Riffusion~\cite{web_reference} exploits the image characteristics of mel-spectrograms and fine-tunes stable diffusion models on a small-scale dataset of aligned music mel-spectrograms and text. This approach achieves impressive results in generating high-quality music guided by text.
\citet{schneider2023archisound} proposes a text-guided latent diffusion method with stacked 1D U-Nets, which can generate multi-minute music from text. Mo\^usai~\cite{schneider2023mo} designs a diffusion model-based audio encoder and decoder to generate high-quality and long-term music from text. Noise2Music~\cite{huang2023noise2music} utilizes Mulan~\cite{huang2022mulan}
and cascade diffusion models to generate high-quality 30-second music clips. MusicLM~\cite{agostinelli2023musiclm} leverages cascade transformers to achieve impressive performance in diverse audio generation tasks. It builds upon the foundations of Mulan~\cite{huang2022mulan} and AudioLM~\cite{borsos2023audiolm}, demonstrating particular proficiency in melody-guided music generation.
MUSICGEN~\cite{copet2023simple} achieves text-conditioned music generation using a single-stage transformer by introducing innovative token interleaving patterns.
These methods utilize large pretrained models to achieve rough music stylization through text, whereas our method can accomplish accurate music style transfer even based on a single example.

\paragraph{Textual inversion.}
While text-guided content generation has achieved impressive results, relying solely on text may not provide precise control over specific aspects, such as editing the style of a piece of music. However, certain works in the field of image generation have explored the potential of textual inversion techniques to personalize the generation process of models.
\citet{gal2022image} propose a textual inversion method that gradually updates the embedding corresponding to the pseudo-word in a pre-trained large language model to represent the visual features of specific objects. There are many variants of this work~\cite{gal2023encoder, li2023stylediffusion, huang2023reversion, tewel2023key, zhang2023inversion, voynov2023p+, zhang2023prospect}.
\citet{zhang2023inversion} uses attention mechanisms~\cite{Guo2023-gw} and CLIP~\cite{radford2021learning} to map images to text embeddings, achieving high-quality image style transfer with a single instance.
ProSpect~\cite{zhang2023prospect} introduces different embeddings to represent the pseudo-word for different generation stages, achieving personalized image generation with the disentanglement of attributes.
Those methods provide us with insights into music style transfer.
\begin{figure*}[ht]
\centering
   \includegraphics[width=\linewidth]{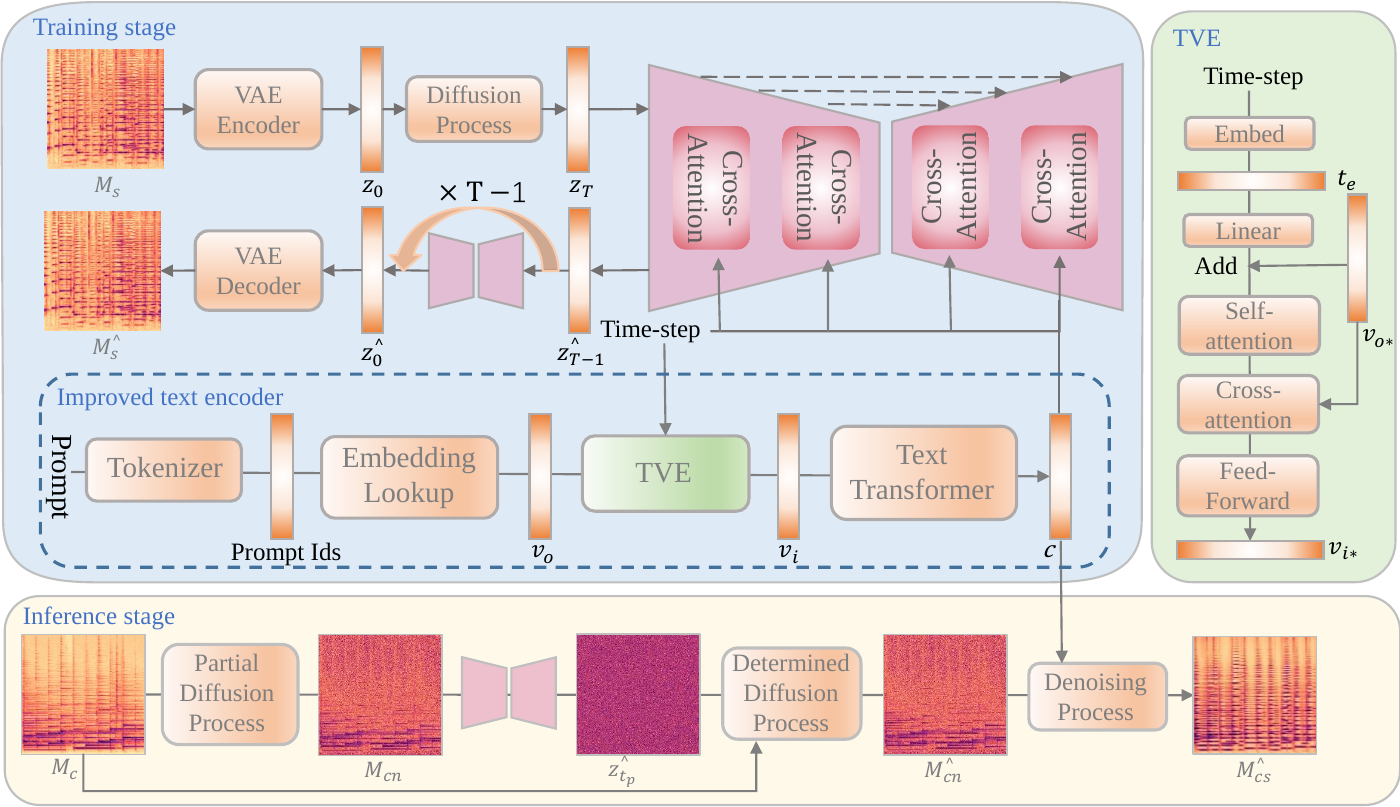}
   \caption{An overview of our method. We adopt Riffusion~\cite {web_reference} as the backbone network and propose a time-varying textual inversion module, which mainly consists of a time-varying encoder (TVE) as shown on the right. 
   Performing several linear layers on the timestep $t_e$, and then adding the output to the initial embedding $v_{o*}$, TVE gives the final embedding $v_{i*}$ through multiple attention modules. 
   $M_s$, $\hat{M}_{s}$, $M_c$, $M_{cn}$, $\hat{z}_{t_p}$, $\hat{M}_{cn}$, $\hat{M}_{cs}$ respectively represent style mel-spectrogram, reconstructed style mel-spectrogram, content mel-spectrogram, noisy content mel-spectrogram, predicted noise, predicted noisy content mel-spectrogam and stylized mel-spectrogram.}
\label{fig:method}
\end{figure*}
\begin{figure*}[!htp]
\centering
   \includegraphics[width=\linewidth]{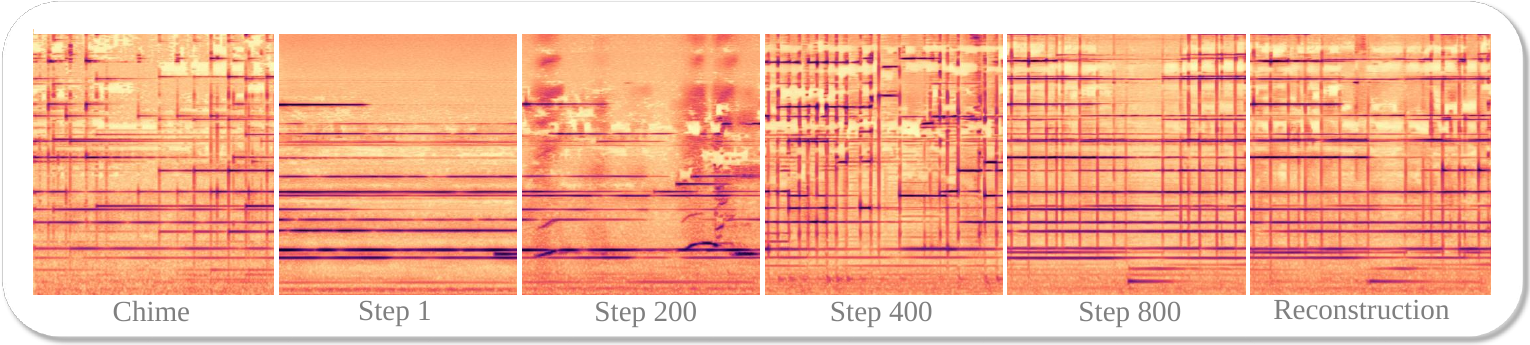}
   \caption{Our time-varying textual inversion module extends the time-step dimension of text embeddings. When reconstructing style mel-spectrograms, the text embeddings exhibit differentiation in the time-step dimension. As the time steps increase, the focus of the text embeddings shifts from texture to structure. }
\label{fig:time_varying}
\end{figure*}

\section*{Method}
We utilize Riffusion~\cite{web_reference} as the backbone to achieve music stylization, as shown in Figure~\ref{fig:method}.
Our work is conducted in the audio frequency domain based on the idea of inversion~\cite{gal2022image}.
During the training stage, we employ our time-varying textual inversion coupled with the diffusion model to iteratively reconstruct the original mel-spectrogram to obtain a pseudo-word representing the style audio.
During inference, guided by the pseudo-word, we incorporate a bias-reduced stylization technique to achieve stable results.

\subsection*{Time-Varying Textual Inversion}
Our approach aims to embed an audio (a piece of music or a natural sound clip) into the latent space of a pre-trained text encoder, obtaining a pseudo-word with text embedding that represents its style.

Latent Diffusion Models (LDMs)~\cite{rombach2022high} take the outputs of the text encoder of CLIP~\cite{radford2021learning} as the condition for text-to-image generation. 
Specifically, the CLIP text encoder tokenizes natural language into multiple indices, each corresponding to an embedding in the embedding lookup. Once the indices are transformed into embeddings $v_o$, they are encoded as conditions for LDMs.In our task, we utilize a pseudo-word ``$*$'' to represent the style audio, which is challenging to express accurately using natural language. The parameters of LDMs are fixed, and the embedding $v_{i*}$ of the placeholder is iteratively updated with the loss of the LDMs until the model can successfully reconstruct the style mel-spectrogram.

The learned ``$*$'' represents the entire style audio, but the structural information (e.g., melody or rhythm) should not be involved in the stylization process. 
By analyzing the diffusion process of the diffusion model, we observe that different timesteps of the diffusion model focus on mel-spectrogram features at different levels. 
We propose a time-varying textual inversion, where the text embeddings of the same pseudo-word change over different timesteps. 
Our experiments show that the text embedding of ``$*$'' exhibits differentiation in the timestep dimension (Figure~\ref{fig:time_varying}).
As the timestep increases, the text embedding gradually focuses more on structure rather than texture. Therefore, we can treat the text embeddings at smaller time steps of the diffusion model as representations of style.

Specifically, we supply timestep $t$ to the time-varying encoder (TVE) module. The timestep is firstly embedded as $t_e$. After performing several linear layers on it, the output is added to the initial embedding $v_{o*}$ as $v^0$, and then undergoes multiple attention modules to derive the final embedding $v_{i*}$. The multiple attention modules start with $v^0$, then each attention layer is implemented as follows:
\begin{equation}
  Attention(Q, K, V) = softmax(\frac{QK^T}{\sqrt{d}}) \cdot V.
\end{equation}
For self attention layer, $Q^s, K^s, V^s$ are defined as:
\begin{equation}
    M^s = W_{M^s} \cdot v^0,
\end{equation}
where $M^s$ can be from $\{Q^s, K^s, V^s\}$.
As for cross attention layer, $Q^c, K^c, V^c$ are defined as:
\begin{equation}
    Q^c = W_{Q^c} \cdot v^1, M^c = W_{M^c} \cdot v^0,
\end{equation}
\begin{equation}
    v^1= Attention(Q^s, K^s, V^s),
\end{equation}
where $M^c$ can be from $\{K^c, V^c\}$.

The final embedding $v_{i*}$ are defined as:
\begin{equation}
    v_{i*}= Attention(Q^c, K^c, V^c).
\end{equation}

By performing text transformer, $v_{i}$ is transformed into conditions for guiding LDMs. Our improved text encoder $e$ is constructed by integrating the CLIP~\cite{radford2021learning} text encoder with TVE.
Based on the loss of LDMs, our optimization objective is defined as follows:
\begin{equation}
    v_{i*}=\underset{v}{\arg\min}\mathbb{E}_{z,y,\epsilon,t}[\|\epsilon-\epsilon_{\theta}(z_{t}, t, e_{\theta}(y,t))\|_{2}^{2}],
\end{equation}
where $z \sim E(x), \epsilon \sim \mathcal{N}(0,1)$, $\epsilon_{\theta}$ and CLIP text encoder of $e_{\theta}$ are frozen during training to maintain the performance of large pretrained models.

\subsection*{Bias-Reduced Stylization}
We observe that for diffusion models, as the timestep decreases during the denoising process from a noisy image to a real image, the primary structure is initially established, followed by the gradual refinement of details. We employ the strength mechanism during the stylization to achieve content guidance. 

Our bias-reduced stylization involves a partial diffusion process, a determined diffusion process, and a denoising process (see Figure~\ref{fig:method}). The partial diffusion process means adding noise to the content mel-spectrogram $M_c$ until the time-step reaches $t_p$, where $t_p = T  \cdot strength$, and $M_c$ is transformed into a noisy mel-spectrogram $M_{cn}$. The determined diffusion process performs a single step denoising on $M_{cn}$, where the predicted noise $\hat{z}_{t_p}$ is used to replace the random noise when performing the diffusion process, resulting in a new noisy content mel-spectrogram $\hat{M}_{cn}$.
This process can be viewed as introducing a bias into the noisy image to counterbalance the impact of model bias.
The denoising process progressively transforms $\hat{M}_{cn}$ into $\hat{M}_{cs}$ by DDIM~\cite{song2020denoising} with a simple prompt ``$*$". Note that both the diffusion process and denoising process are performed in the latent space of the VAE encoder. The denoised output requires decoding by the VAE decoder into a Mel-spectrogram, which can subsequently be reconstructed into audio using the Griffin-Lim algorithm.
\section*{Experienment}
\begin{figure*}[t]
\centering
   \includegraphics[width=\linewidth]{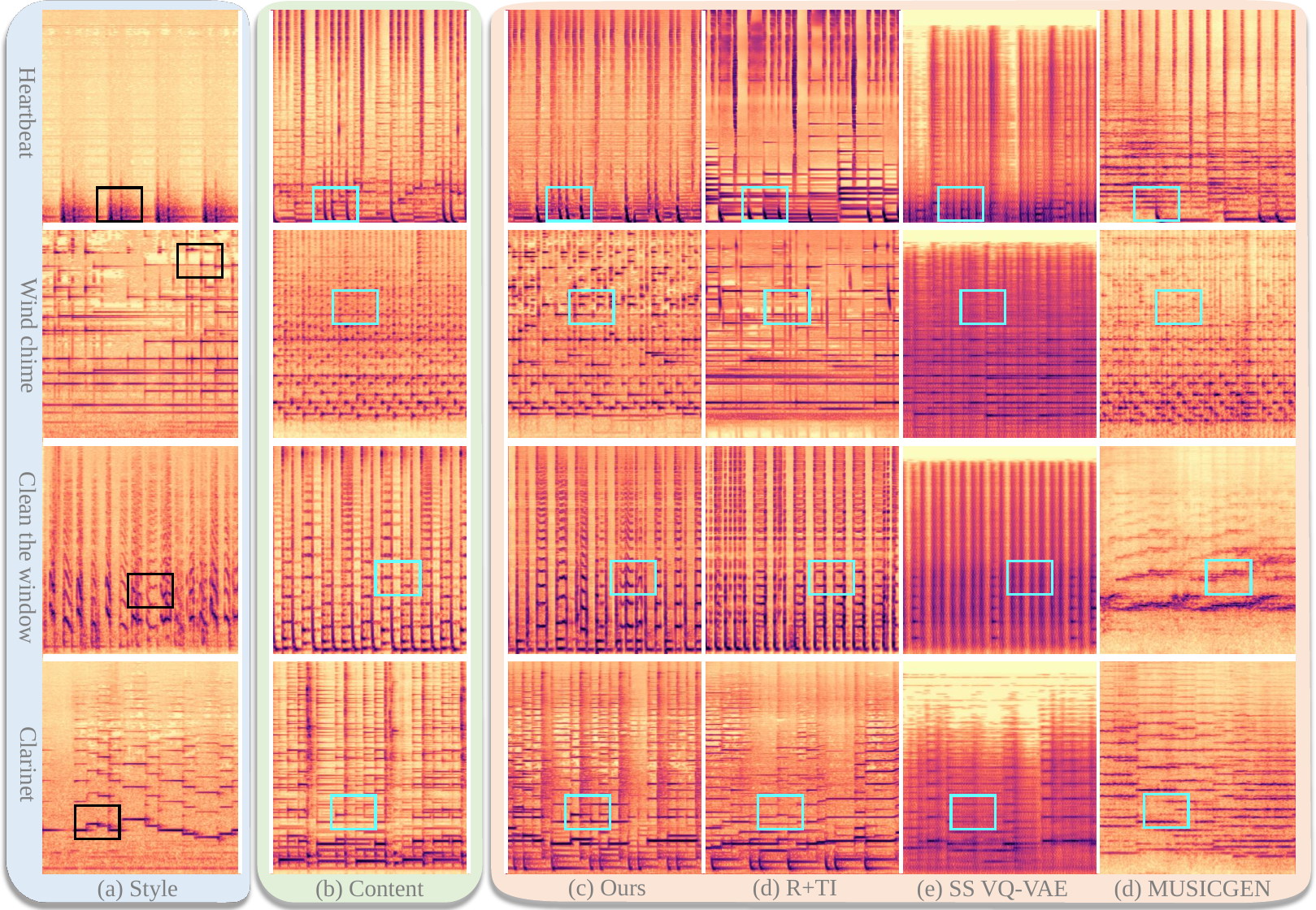}
   \caption{Qualitative comparison with state-of-the-arts methods~\cite{ web_reference, gal2022image, cifka2021self, copet2023simple}. (a) Style mel-spectrograms, the texts on the left are the sound categories. (b) Mel-spectrograms. (c)-(d) The stylized results of various methods. In the style mel-spectrograms, the black box highlights the regions with prominent texture. It can be observed in the blue boxes that only our results preserve a similar structure to the content mel-spectrograms while exhibiting a similar texture to the style mel-spectrograms.}
\label{fig:contrast}
\end{figure*}

We conducted qualitative evaluation, quantitative evaluation and ablation study to demonstrate the effectiveness of our method, which performs well in both content preservation and style fit.
 
\paragraph{Dataset.}
Currently, there is a lack of publicly available datasets specifically tailored for music style transfer that meet our requirements. We collected a small-scale dataset from a website (\url{https://pixabay.com}) where all the content is free for use. 
The collected data was segmented into five-second clips, resulting in a total of 253 5-second clips, with 74 style clips and 179 content clips. The style subset consists of 18 different style audios, including instruments, natural sounds, and synthesized sound effects.
The content subset consists of electronic music and instrument clips, distinguishing it from other music style transfer approaches that primarily employ simple monophonic audio. In our experiments, we did not utilize all of the style audio clips. Instead, we selected only one sample for each natural sound and synthetic sound effect. Considering the variability of musical instrument notes, we used 3-5 clips for each instrument.

We compared our method with three related state-of-the-art approaches:
\begin{itemize}
\item{R+TI: We combined Riffusion (R)~\cite{web_reference} with Textual Inversion (TI)~\cite{gal2022image} as our baseline. R is the original stable diffusion model v1.5, which is just fine-tuned on images of mel-spectrograms paired with text. Additionally, it incorporates a conversion library for transformation between audio and mel-spectrograms. TI is a classical method that learns a pseudo-word for a concept within a limited number of images using an optimization-based approach.}
\item{SS VQ-VAE~\cite{cifka2021self}: A  latest available implementation of one-shot music style transfer.}
\item{MUSICGEN~\cite{copet2023simple}: A recently released text-guided music generation method that achieves text-guided music stylization with melody conditioning.}
\end{itemize}

\paragraph{Implementation details.}
In our experiments, we fix the parameters of LDMs and text encoder except for the TVE module. We use the default hyperparameters of LDMs and set a base learning rate of 0.001. The training process on each style takes approximately 30 minutes using an NVIDIA GeForce RTX3090 with a batch size of 1, less than the more than 60 minutes required for TI. During 
inference, our approach employs two hyperparameters: $strength$ and $scale$. These parameters respectively govern the intensity of the content and regulate the intensity of the style. We achieved the best results when strength ranged from 0.6 to 0.7 and the scale ranged from 3.0 to 5.0.\

\subsection*{Qualitative Evaluation}
The stylized audio samples, showcasing the comparison between our method and other approaches, can be accessed on the static webpage provided within the supplementary materials.
As shown in the Figure~\ref{fig:contrast}, we compared our method with three approaches: R+TI~\cite{web_reference, gal2022image}, SS VQ-VAE~\cite{cifka2021self}, and MUSICGEN~\cite{copet2023simple}. The structure of the mel-spectrogram can be seen as the content, while the detailed texture is considered as the style. 

For R+TI, we treated partial noisy content mel-spectrogram as content guidance and used the learned pseudo-word as text guidance for style transfer using DDIM. It can be observed that although R+TI preserves the overall structure well, it introduces occasional flaws in the rhythm at the local level and exhibits weaker texture transfer compared to our method.
SS VQ-VAE processes audios with a sampling rate of 16kHz, resulting in the loss of high-frequency information after stylization. It introduces severe artifacts in the mel-spectrogram, resulting in poor performance in terms of audio quality.
Regarding MUSICGEN, we used the textual descriptions of the style audios as guidance for style transfer. The results indicate that its generation quality exhibits a high degree of stochasticity, characterized by unstable content preservation and limited editability.   
Our method can accurately preserve the structure of content mel-spectrograms while achieving high-quality texture transfer of style mel-spectrograms, without introducing the artifacts observed in other methods.

\subsection*{Quantitative Evaluation}
Following the previous works on music style transfer~\cite{alinoori2022music, cifka2021self}, we evaluate our method based on two criteria: (a) content preservation and (b) style fit. Taking inspiration from MUSICGEN~\cite{copet2023simple} and InST~\cite{zhang2023inversion}, we compute the CLAP cosine similarity between the generated mel-spectrograms and the content mel-spectrograms to evaluate content preservation. Additionally, we calculate the CLAP cosine similarity between the generated mel-spectrograms and the corresponding textual description of the style to evaluate style fit. 
We computed the CLAP cosine similarity between the textual descriptions and the style mel-spectrograms as a reference, with an average value of 0.4890 and a minimum value of 0.3424.
Thus, we excluded style audios that were difficult to describe in text from the calculation of objective metrics.
This ensures the correlation between our style mel-spectrograms and the evaluation text.
We evaluated our method and other approaches by randomly selecting 282 content-style pairs and assessing their performance, as shown in Table~\ref{tab:contrast}. Our method achieves the best performance in both metrics, significantly surpassing our baseline in terms of content preservation. While SS VQ-VAE achieves a similar style fit to ours, it suffers from greater content loss. MUSICGEN performs noticeably worse than our method in both metrics.
\begin{table}
   \begin{center}
   \begin{tabular}{c|c|c|c|c|c}
   \hline
  {}&\multicolumn{2}{|c}{Objective} & \multicolumn{3}{|c}{Subjective}\\
    Method & CP & SF & CP & SF & OVL \\ 
   \hline
   R+TI & 0.3481 & 0.2722 & 2.81 & 3.20 &2.75\\
   SS VQ-VAE & 0.2351 & 0.2809 & 3.36 & 2.34 &2.60\\
   MUSICGEN & 0.2808 & 0.2370 & 2.81 & 2.70&2.83\\
   \textbf{Ours}& \textbf{0.4645} & \textbf{0.2816} & \textbf{3.91} & \textbf{3.70}&\textbf{3.66}\\
   \hline
   \end{tabular}
   \end{center}
   \caption{Qualitative comparison with other methods~\cite{ web_reference, gal2022image, cifka2021self, copet2023simple}. CP, SF, OVL stands for Content Preservation, Style Fit, and Overall Quality, respectively. }
   \label{tab:contrast}
\end{table}

\paragraph{User study.}
To conduct a subjective evaluation of our method's performance, we designed a user study to rate the four methods on three evaluation metrics. We randomly selected 15 sets of results (excluding comparisons with MUSICGEN~\cite{copet2023simple} for style audios that are difficult to describe with text). Before the test, we set up questions to assess the participants' music profession level and provided guidelines outlining the evaluation criteria for music style transfer. During the test, each participant was presented with a style audio, a content audio, and four randomly ordered generation results for each 
set of questions.
Participants were asked to rate the following metrics on a scale of 1 (lowest) to 5 (highest):
\begin{itemize}
    \item Content Preservation: consistency between the generated audio and the content music in terms of melody, rhythm, and similar attributes.
    \item Style Fit: consistency between the generated audio and the style audio in terms of timbre, sound units, and similar attributes.
    \item Overall Quality: the quality related to the overall performance of style transfer, such as the coherence of the fusion between the content and style of generated music.
\end{itemize}
Our experiment involved 80 participants, out of which 72 were deemed valid (excluding participants with no knowledge of music), resulting in a total of 12960 ratings. After excluding the maximum and minimum values, We calculated the weighted average based on participants' music profession level (four levels with corresponding weights: 1 to 4). The results, as presented in Table~\ref{tab:contrast}, demonstrate that our method outperforms other approaches significantly in terms of content preservation, style 
fit, and overall quality.

\subsection*{Abaltion Study}
\begin{table}
   \begin{center}
   \begin{tabular}{c|c|c}
   \hline
    {} & Content Preservation & Style Fit  \\
   \hline
   w/o TVE & 0.4506 & 0.2418\\
   w/o BRS & 0.4415 & 0.2602\\
   \textbf{Ours}& \textbf{0.4645} & \textbf{0.2816}\\
   \hline
   \end{tabular}
   \end{center}
    \caption{Ablation study of our method. TVE and BR are Time-Varying Embedding and Bias-Reduced Stylization respectively.}
   \label{tab:ablation}
\end{table}

\paragraph{Time-varying embedding (TVE).}
We fix the text embedding of the pseudo-word at a specific time step during inference and use it as the text guidance for mel-spectrogram generation, as shown in Figure~\ref{fig:time_varying}. As the timestep increases, the text embeddings gradually shift their focus from the texture of the mel-spectrogram to the structure. This aligns with our expectation that the diffusion model first constructs the 
rough structure of the image during denoising and then optimizes the details. The reconstructed results reflect the high-quality reconstruction due to the fusion of features across different timesteps. To further demonstrate the effectiveness of the TVE module, we evaluate our method without it, as shown in Table~\ref{tab:ablation}. Although the difference in content preservation is not significant after removing TVE, there is a noticeable decrease in style fit, indicating that TVE contributes to better style learning.

\paragraph{Bias-reduced stylization.}
We evaluate the impact of removing the bias-reduced stylization technique on content preservation and style matching. It can be observed that there is a decrease in both metrics, indicating that it is helpful in terms of preserving content and facilitating style transfer.

\subsection*{Discussions and Limitations}
Our method enables music style transfer using diverse audio sources, including instruments, natural sounds, and synthesized sound effects. Nevertheless, it is crucial to recognize that certain limitations may arise in specific contexts. For instance, when the content music encompasses multiple components, our method may encounter challenges in accurately performing style transfer on each individual component, potentially leading to partial content loss. Furthermore, when the style audio incorporates white noise like rain or wind sounds, it becomes challenging to capture the inherent musicality within those elements and transfer it effectively to the content reference.
\section*{Conclusion}
In this paper, we propose a novel approach for music stylization based on diffusion models and time-varying textual inversion, which effectively embeds style mel-spectrograms. Our experiments demonstrate the generality of our method for various types of audio, including musical instruments, natural sounds, and synthesized sound effects. Our approach achieves style transfer with a small amount of data, generating highly creative music. Even when applied to non-musical style audio, our method produces results with a high level of musicality. We believe that leveraging pre-trained models with stronger generative capabilities would further enhance the performance of our method. In the future, we aim to investigate more interpretable and attribute-disentangled music style transfer.
\section*{Acknowledgements}
This work was supported by the National Natural Science Foundation of China under nos. 61832016 and 62102162.

\bibliography{DiffMusicST}

\end{document}